\documentclass[a4paper,11pt]{article}
\pdfoutput=1 

\usepackage{jheppub} 

\usepackage[T1]{fontenc} 

\title{The Wilson Flow and the finite temperature phase transition}

\author[a]{M. Wandelt,}
\author[a]{F. Knechtli,}
\author[a]{M. G\"unther}

\affiliation[a]{Bergische Universit\"at Wuppertal,\\Gau{\ss}str. 20, 42119 Wuppertal, Germany}

\emailAdd{wandelt@math.uni-wuppertal.de}
\emailAdd{guenther@math.uni-wuppertal.de}
\emailAdd{knechtli@physik.uni-wuppertal.de}

\abstract{
We consider the determination of the finite temperature phase transition in the 
Yang--Mills SU(3) gauge theory.
We compute the difference of the spatial and temporal energy density at a 
physical Wilson flow time. This difference is zero in the confined phase and
becomes non zero in the deconfined phase. We locate the phase transition
by using a new technique based on an exponential smoothing spline.
This method is an alternative to the determination of the phase transition
based on the Polyakov loop susceptibility and can also be used with dynamical 
fermions.
}

\begin{document} 
\maketitle
\flushbottom

\section{Introduction}
\label{sec:intro}

The pure SU(3) gauge theory at finite temperature exhibits a first order
phase transition separating the confined phase at low temperature from the
deconfined phase at high temperature. The standard way to locate the 
critical temperature by Monte Carlo simulations is based on the computation
of the susceptibility of the Polyakov loop. The position of the peak
of the susceptibility determines the
temperature and the scaling of the peak value with the spatial volume identifies
the order of the transition.

In this work we study an alternative method to locate the phase transition
based on the energy density computed using the Wilson flow. We separate
the spatial and temporal components of the energy density. Their difference
is zero in the confined phase and becomes non zero in the deconfined phase.
This behaviour allows to locate the phase transition. Technically we propose
to use an exponential smoothing spline to fit the data of the energy difference.
We give explicit formulae for the construction of such a spline. We
define the critical temperature to correspond to the maximum slope of the
spline. We verify by Monte Carlo simulations that the new method agrees with
the results of the standard method based on the Polyakov loop susceptibility.

The motivation to look at an alternative method to locate the phase transition
is that in QCD with dynamical fermions one usually uses the peak of the chiral 
susceptibility to locate the phase transition or cross over (depending on the quark mass), which is a computationally 
expensive quantity. The method based on the energy difference which we describe
in this paper can be applied also in presence of dynamical fermions.

\section{Finite temperature phase transition}
\label{sec:FiniteTemperature}

The finite temperature can be investigated on lattices where the time extension $N_t$ is much smaller than the spatial one $N_s$. 
This means, it should at least hold $N_s \geq 4 \cdot N_t$ as described in \cite{Boyd}. 
The temperature $T$ is related to the time extension by the equation $$T(\beta)=\frac{1} {N_t \cdot a(\beta)}$$
with lattice spacing $a(\beta)$.
This implies for a given $N_t$, there is a critical coupling $\beta_c$ such that the critical temperature can be computed via $T_c(\beta_c).$
Finally, the critical temperature can be computed in physical units, for example, as 
\begin{align}\label{r0}
T_c [\text{MeV}] &= \frac{r_0/a(\beta_c)}{N_t \cdot r_0} \cdot \hbar c =  \frac{r_0/a(\beta_c)}{N_t \cdot 0.49} \cdot 197.3 \; \text{MeV} 
\end{align}
with $\hbar c=197.3$ MeV fm and $r_0=0.49$ fm \cite{r0}. The formula for the determination of the scale $r_0/a(\beta)$ is 
\begin{equation}\label{eq:ln:a:r0}
  \ln (a/r_0) = -1.6804 - 1.7331 (\beta-6) + 0.7849 (\beta-6)^2 - 0.4428 (\beta-6)^3
\end{equation}
for $5.7 \leq \beta \leq 6.92$
and explained in \cite{Necco:Sommer:2002}.
\subsection{Polyakov loop susceptibility}\label{subsec:PLS}
\begin{figure}[h]\begin{center}
    \includegraphics[width=10cm]{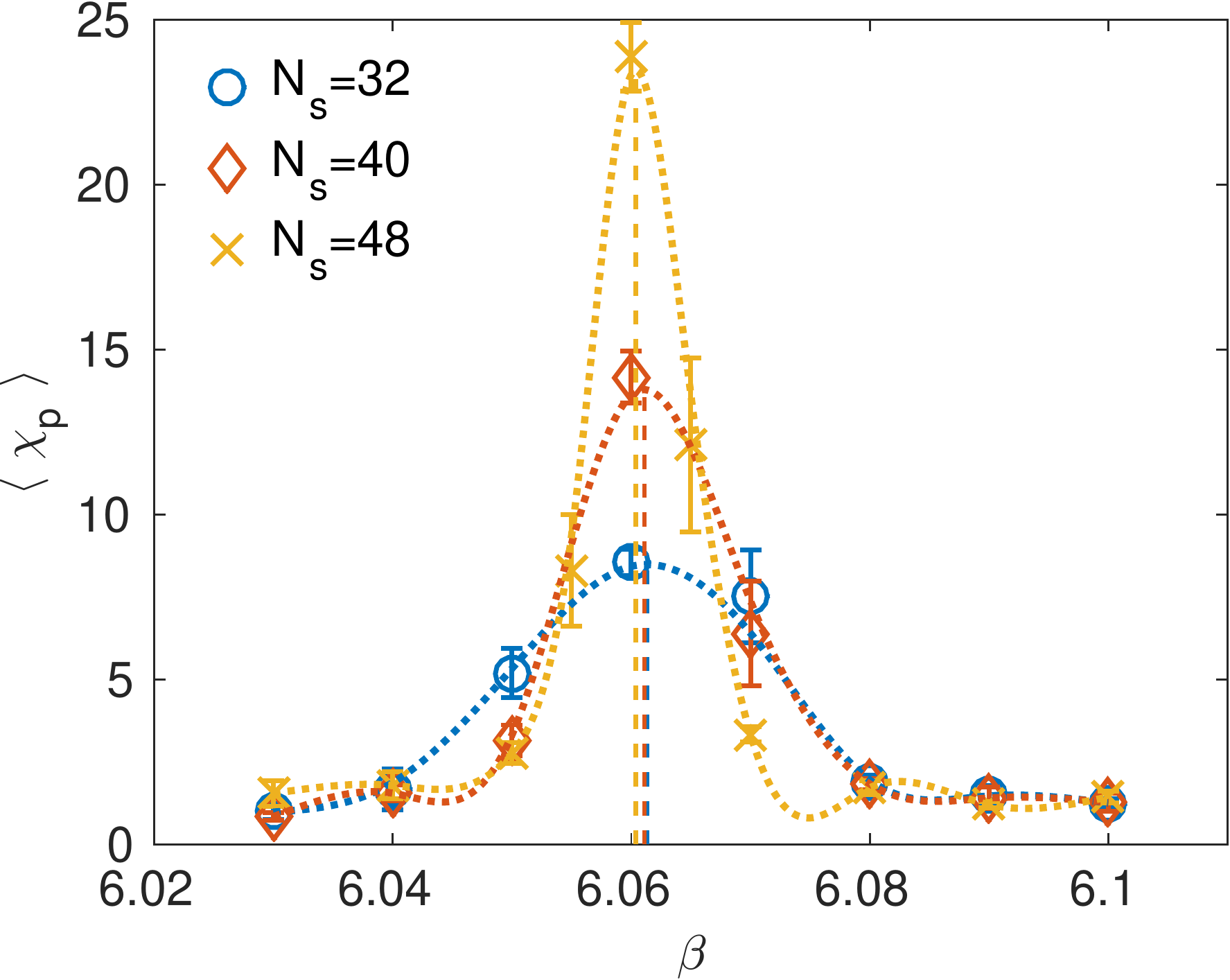}
    \caption{Standard approach: detection of the critical coupling using the Polyakov loop susceptibility, here for $N_t=8$. 
    The data are fitted by an exponential smoothing spline and the critical coupling is determined via its maximum. 
   }
    \label{fig:polyakov}
  \end{center} \end{figure}

Usually, the critical coupling $\beta_c$ is determined using the Polyakov loop susceptibility.
The lattice average $P$ of the Polyakov loop is computed as
    $$ P= \frac{1}{N_s^3} \sum_{\vec{x}} \frac13 \text{Tr} \prod_{n_t=1}^{N_t} U(\vec{x},n_t)$$ 
which is the mean of the product of links $U(\vec{x},n_t)$ in time direction $n_t$ for each 3d-spatial vector $\vec{x}$. 
As the lattice average $P$ is a complex number, the Polyakov loop susceptibility $\chi_p$ is the variance of its absolute value $|P|$,
$$\chi_p := N_s^3 \cdot \Bigl(\langle |P|^2\rangle - \langle |P|\rangle^2 \Bigr).$$
For a given lattice of size $(N_t,N_s^3)$, the coupling constant $\beta$ has to be chosen and through Monte Carlo simulation an ensemble of gauge configurations
is generated. The expectation value in the ensemble
of the Polyakov loop susceptibility $\langle \chi_p(\beta) \rangle$ can be computed.
This has to be done for several values of $\beta$ around the critical coupling $\beta_c$ such that the curve $(\beta, \langle \chi_p \rangle)_k$ for $k=1,\ldots,n_\beta$ contains a peak around $\beta_c$. The desired value $\beta_c$ is found by a fit of the data, for example, with a smoothing spline $s(\beta)$.
The critical coupling $\beta_c$ is computed as the value of $\beta$ at which the fit $s(\beta)$ reaches its maximum value.
Fixing $N_t$ and varying $N_s$ leads to several values $\beta_c(\chi_p, N_t, N_s)$ of the critical coupling. Figure \ref{fig:polyakov} shows the results of
the simulations which we will describe in section \ref{sec:results}.
At the end, a linear extrapolation in $1/N_s^3$ leads to the desired critical coupling $\beta_c(\chi_p, N_t,\infty)$ for infinite volume.

\subsection{The Wilson flow}\label{subsec:WF}
The Wilson flow \cite{Narayanan:Neuberger:2006,Luescher:triv,Lohmeyer:Neuberger:2011} $V(t)$
is a flow of lattice gauge fields and it is defined as the solution of the ordinary Lie group differential equation
\begin{align}\label{eq:wilson_flow}
  \dot{V}_{x,\mu}(t) = Z_{x,\mu}\bigl(V(t)\bigr) \cdot V_{x,\mu}(t)
\end{align}
with link variables $V_{x,\mu}(t)$ being elements of the special unitary Lie group $\text{SU}(3)$ 
and a function $Z_{x,\mu}(V(t))$ which takes values in the Lie algebra $\mathfrak{su}(3)$.
Particularly, the function $Z_{x,\mu}$ is the Lie derivative $Z_{x,\mu}\bigl(V(t)\bigr) = -\partial / \partial V_{x,\mu}(t) S_W$ of the Wilson action
$$ S_W = \sum_{p} \text{Re} \text{Tr} \{1-V_{p}(t)\}$$
such that it depends not only on the link itself but also on its staples. 
$S_W$ is the sum over all oriented plaquettes $p$ and $V_{p}(t)$ is the product of link variables around one of these plaquettes $p$, see \cite{Luescher:prop}. 
The Wilson flow \eqref{eq:wilson_flow} can be integrated numerically with initial values $V(t_0)$ taken from an ensemble of configurations generated for a particular value of the coupling constant $\beta$.
The numerical integration is performed up to a certain flow time and during this computation gauge invariant observables of interest are measured. 
Thereby, we focus on the energy density 
\begin{align}\label{eq:energy_density}
 E &= \frac{a^4}{4} G_{\mu \nu}^c G_{\mu \nu}^c
\end{align}
with lattice spacing $a$ and symmetric field strength tensor $G_{\mu \nu}$ as described in \cite{Luescher:prop}.

\subsection{Difference in the Wilson energy}\label{subsec:Difference}
We investigate the mean energy density $\langle E \rangle$ described in equation \eqref{eq:energy_density} on a four-dimensional lattice with fixed temporal lattice size $N_t=8$. 
Particularly, we focus on the mean spatial and temporal part of the energy $\langle E_{ss}\rangle$ and $\langle E_{st}\rangle$ and 
its difference $\langle \Delta E \rangle:= \langle E_{ss}-E_{st}\rangle$.

\begin{figure}\begin{center}
  \includegraphics[width=10cm]{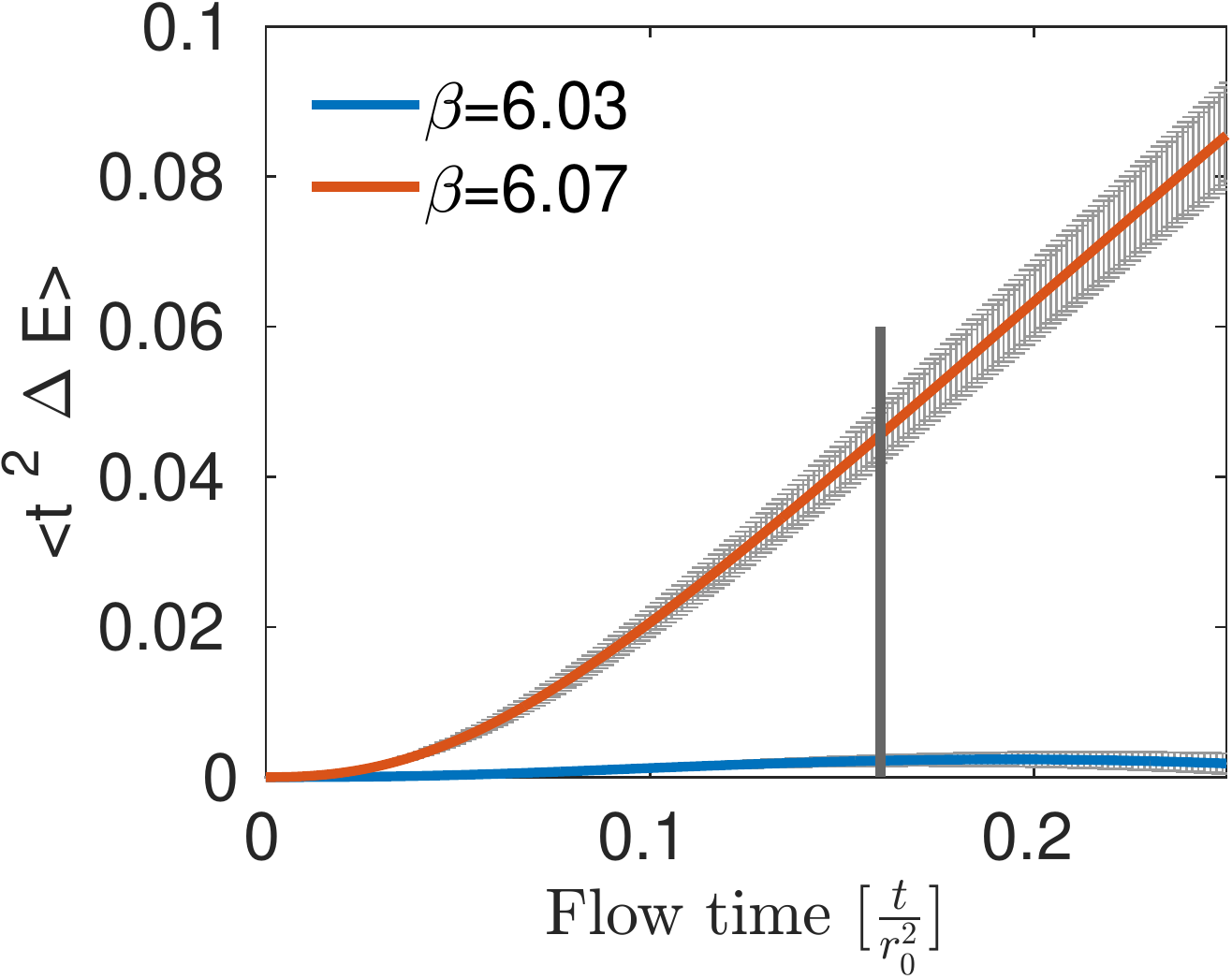}
  \caption{Difference in the temporal and spatial energy for values of $\beta=6.03$ and $\beta=6.07$. 
  In the confined phase (blue) with values of $\beta < \beta_c$ the difference $\Delta E$ in the spatial and the temporal part of the energy is approximately $0$, 
  $\Delta E \gg 0$ in the deconfined phase (red).}
  \label{fig:Ediff}
\end{center} \end{figure}

The splitting in the temporal and spatial part is done as follows numbering the dimensions $0,1,2,3$ with $0$ being the time dimension: the spatial planes are $(1,2), (1,3)$, and $(2,3)$ and the temporal ones $(0,k)$ with $k=1,2,3$.
Here, the exact formulae for the spatial part of the energy $E_{ss}$ and the temporal one $E_{st}$ are 
$$E_{st} = \frac{a^4}{4} G_{0 \nu}^c G_{0 \nu}^c \quad \text{and}\quad E_{ss} = \frac{a^4}{4} G_{\mu \nu}^c G_{\mu \nu}^c \quad \text{with} \;\;{\mu,\nu=1,2,3},\;\mu < \nu.$$
The critical coupling $\beta_c$ divides the confined phase and the deconfined phase. 
In the confined phase, the values of $\beta$ are smaller than $\beta_c$, in the deconfined one larger. 
It is known from \cite{Boyd} that the critical coupling for lattices with $N_t=8$ is at $\beta_c$=6.0625. 
For a first test, we simulated lattices with $\beta=6.03<\beta_c$ and $\beta=6.07>\beta_c$ and computed $\langle E_{ss}\rangle$, $\langle E_{st}\rangle$ and its difference $\langle \Delta E \rangle$ as a function of the flow time. 
It can be seen in figure \ref{fig:Ediff} that in the confined phase (blue) with values of $\beta$ smaller than $\beta_c$ there is no difference in the spatial and the temporal part of the energy. 
On the other hand, in the deconfined phase (red) with values of $\beta$ larger than $\beta_c$, there is a difference in both parts of the energy.
This implies, the spatial and temporal mean energy densities $\langle E_{ss}\rangle$ and $\langle E_{st}\rangle$ coincide in the confined phase and 
differ in the deconfined one.
In the following paragraphs, 
we fix a certain flow time $\sqrt{t}/r_0$ in units of the scale $r_0$ such that $\langle \Delta E \rangle$ means
$$\langle \Delta E(\sqrt{t}/r_0) \rangle := \langle E_{ss}(\sqrt{t}/r_0) - E_{st}(\sqrt{t}/r_0)\rangle\;.$$
For a fixed flow time $\sqrt{t}/r_0$, it is shown in figure \ref{fig:Ediff} (in this case $\sqrt{t}/r_0=0.4$)
that the energy difference $\Delta E$ is approximately zero in the confined phase and grows suddenly in the deconfined one. This advises that the critical coupling can be found using the energy difference at a certain flow time.
 
\subsection{The energy difference method}\label{subsec:EDM}
\begin{figure}[t]\begin{center}
    \includegraphics[width=10cm]{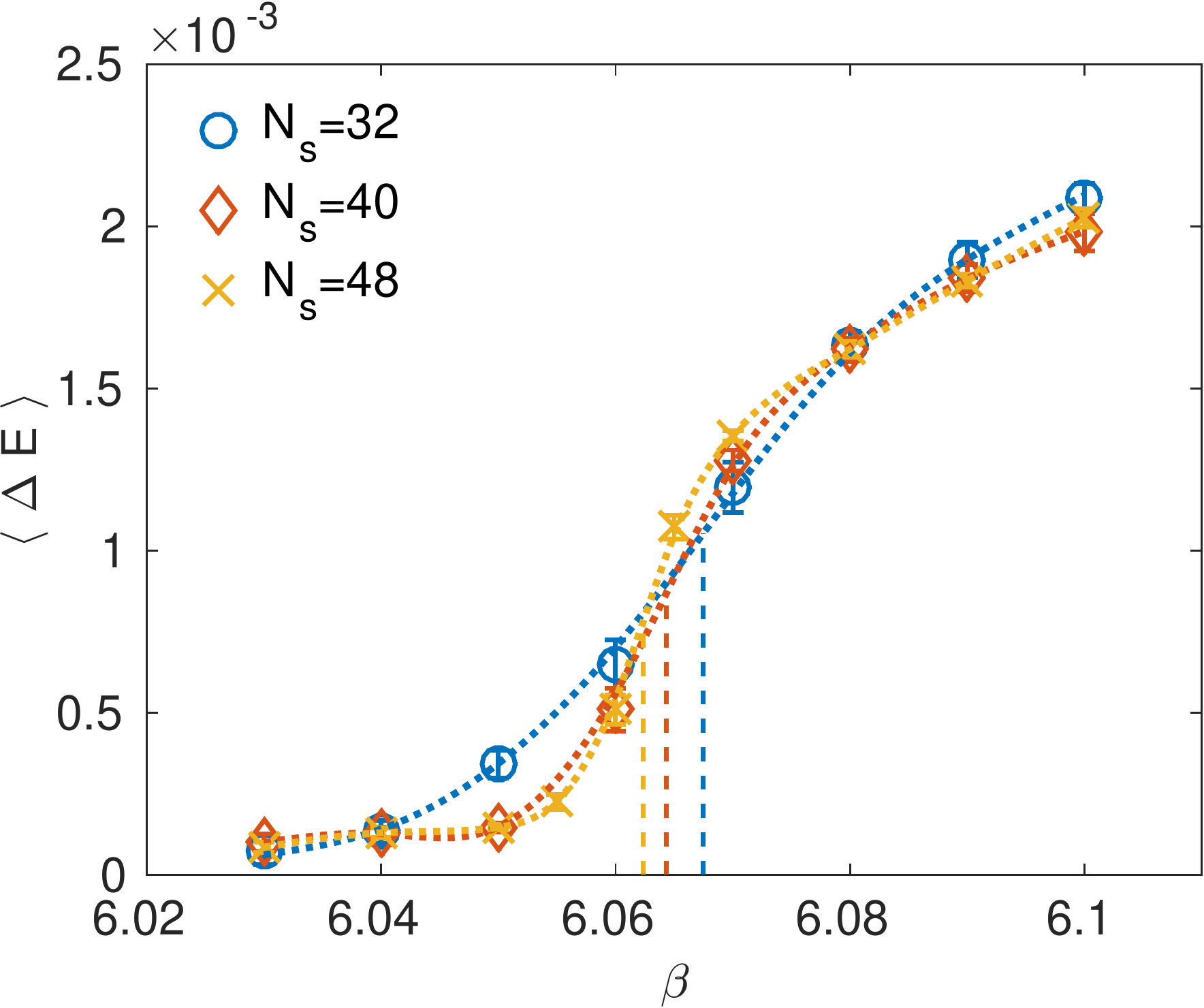}
    \caption{Detection of the critical coupling using the energy difference method. The data are fitted by an exponential smoothing spline. For each lattice, the critical coupling is computed as value of $\beta$ at which the maximum slope of the spline occurs.
     }
   \label{fig:energyDiff}
\end{center} \end{figure}

We developed a new method\footnote{\label{foot:1}In the course of our work the article \cite{Datta} appeared, which also uses this method.} for the detection of the critical temperature and called it the energy difference method. 
Therefore, we consider the Wilson flow and focus on the difference in the spatial and temporal energy density at a certain flow time $\sqrt{t}/r_0$. 
The critical coupling $\beta_c$ can be computed by a fit through an exponential smoothing spline developed in \cite{WGW}. 
The idea is to combine smoothing splines, which approximate the data in a spline setting, with an exponential spline, such that undesired oscillations not given in the data are avoided (see section \ref{sec:ESS} for more details). 
For the detection of the critical coupling, we proceed as follows similarly to the standard Polyakov loop susceptibility approach:
first, we fix a lattice size $(N_t,N_s)$ and the values of $\beta_k$ for $k=1,\ldots,n_\beta$. Then, we need the results of a HMC or heat bath simulation for these values as input for the Wilson flow.
Moreover, the Wilson flow has to be computed up to a certain simulation time which has to be determined such that the statistical errors are minimized. 
We fixed the time $\sqrt{t}/r_0=0.15$ as described in paragraph \ref{subsection:stat_err}. Additionally, the values for the spatial and temporal energy density have to be measured for the flow time $\sqrt{t}/r_0=0.15$.
After the simulation, an exponential smoothing spline $s(\beta)$ is determined to fit the data $(\beta,\langle \Delta E\rangle)$.
The critical coupling $\beta_c(\Delta E, N_t, N_s)$ for the specific lattice size $(N_t, N_s^3)$ is determined as value of $\beta$ where the steepest gradient of the spline $s(\beta)$ occurs.
This has to be repeated for a few spatial lattice sizes $N_s$ such that the values $\beta_c(\Delta E, N_t, N_s)$ can be extrapolated towards an infinite space dimension $N_s=\infty$
in order to compute $\beta_c(\Delta E, N_t, \infty)$ for the finite temperature phase transition in infinite volume. Figure \ref{fig:energyDiff}
shows the results of
the simulations which we will describe in section \ref{sec:results}.

The energy difference $\Delta E$ is equivalent (up to discretization effects)
to the difference of the temporal and spatial plaquettes, cf. \cite{Luescher:prop}, and it corresponds 
to the sum of the pressure and the energy density, see e.g. \cite{Engels1994}.
It is expected to develop a discontinuity in the thermodynamic limit
\footnote{
We thank the referee of our paper for pointing this out.} since the
energy density is discontinuos in the Yang--Mills SU(3) theory, see e.g. \cite{Boyd}.
Presently our data shown in Figure \ref{fig:energyDiff} do not allow to
verify this expectation.

\section{Exponential smoothing splines}\label{sec:ESS}
The class of exponential smoothing splines approximates data with uncertainties avoiding undesired oscillations.
It couples the ideas of smoothing splines \cite{Reinsch1967} with exponential splines \cite{Schweikert1966, Rentrop}. 
Here, we explain the idea of exponential smoothing splines and give the necessary formulae for its implementation. 
The mathematical essentials can be found in \cite{WGW} based on \cite{Master:Werneburg} .

\subsection{Idea}\label{subsec:Ess_idea}
We start with data $(x_i,y_i)$ with errors $w_i$, $i=1,\ldots,n$. 
The data should be approximated by a spline $s(x)$ within the region of their errors and, moreover, no artificial oscillations should be added.
The spline can be found by minimizing the energy function
$$\int_{x_1}^{x_n} \left[ f^{\prime\prime}(x)^2 + \Lambda(x)^2 f^\prime(x)^2 \right] dx$$
 among all $f \in C^2(x_1,x_n)$ using the constraints $$\sum_{i=1}^n \biggl(\frac{f(x_i)-y_i}{w_i}\biggr)^2 \leq S$$ 
and piecewise constant tension functions $\Lambda(x)=\lambda_i > 0$ with $x \in[x_{i},x_{i+1})$ for $i=1,\ldots,n$.
Here, $\Lambda(x)$ avoid undesired oscillations, 
the weights $w_i$ are correlated to the errors of $y_i$ 
and the smoothing parameter $S$ defines how much the approximated values $f(x_i)$ may differ from $y_i$.
There are three limiting cases which coincide with already known kinds of splines:
\begin{itemize}
 \item $\lambda =0, S=0$ describes the well-known cubic spline, 
 \item $\lambda =0, S\neq 0$ leads to the smoothing spline which \emph{approximates} the data; it may lead to oscillations not given in the data,
 \item and $\lambda \neq 0, S=0$ results in the exponential spline (also known as spline under tension) \emph{interpolating} the data exactly avoiding oscillations not given in the data.
\end{itemize}
  
\subsection{The shape of the exponential smoothing spline}  
Finally, the exponential smoothing spline is parameterized as
\begin{align}\label{eq:shape_expsmooth}
  s(x)&=s_{i+1} t + s_i (1-t) + \frac{d_{i+1}}{\lambda_i^2} \biggl(\frac{\sinh(\mu_i t)}{\sinh \mu_i} -t\biggr)
  + \frac{d_i}{\lambda_i^2} \biggl(\frac{\sinh(\mu_i(1-t))}{\sinh(\mu_i)}-(1-t)\biggr)\
\end{align}
for $x \in [x_i,x_{i+1})$ and coefficients $s_i,s_{i+1}, d_i, d_{i+1}, \mu_i$ and $\lambda_i$ for all intervals $i=1,\ldots,n-1$. 
For convenience, the variables $h_i$, $t$ and $\mu_i$ are used. They are specified as $$h_i:=x_{i+1}-x_i, \quad t:=\frac{x-x_i}{h_i} \quad \text{and} \quad \mu_i:=h_i\cdot \lambda_i \quad \text{for} \quad i=1,\ldots,n-1.$$
The variables $\lambda_i$, are tension parameters for all intervals $i=1,\ldots,n-1$ and, assuming they are known, $s=(s_1,\ldots,s_n)^T$ and $d=(d_1,\ldots,d_n)^T$ are the only unknowns in equation \eqref{eq:shape_expsmooth}.
\subsection{Linear equations for the unknowns}
So, for given data $(x_i,y_i)$, $i=1,\ldots,n$, the coefficients $s_i$ and $d_i$ can be computed  
via the linear equations
\begin{equation}\label{eq:lin}
Q^Ts=T\tilde{d}\quad \text{and} \quad Us-Q\tilde{d}=pD^{-2}(s-y)
\end{equation}
with $s=(s_1,\ldots,s_n)^T$, $\tilde{d}=(d_2,\ldots,d_{n-1})^T$, $d=(d_1=0,\tilde{d},d_n=0)$, $y=(y_1,\ldots,y_n)^T$ and Lagrange parameter $p$.
The matrices $Q,T,U$ and $D$ are sparse matrices and they have to be set up as follows: 
choose
\begin{align*}
 u_i&:=\frac{\lambda_i^2}{h_i}, & t_i&:=\frac{\cosh(\mu_i)}{\lambda_i \sinh(\mu_i)} - \frac{1}{\lambda_i^2 h_i},\\
 v_j&:=u_j+u_{j+1}, & t_{k,k+1}&:= \frac{1}{\lambda_j^2 h_k} - \frac{1}{\lambda_k \sinh(\mu_k)},\\
 \text{and}\quad q_i&:=\frac{1}{h_i} \\
\end{align*}
for $i=1,\ldots,n-1$, $j=1,\ldots,n-2$ and $k=2,\ldots,n-2$ such that the matrices read
\begin{align*}
Q&:=  \begin{pmatrix}
      -q_1 & & 0 \\
      q_1+q_2 & \ddots& \\
      -q_2 & \ddots & -q_{n-2}\\
      & \ddots & q_{n-2}+q_{n-1}\\
      0 & & -q_{n-1}
     \end{pmatrix},    
&
T&:=\begin{pmatrix}
     t_1+t_2 & t_{23} & & 0\\
     t_{23} & \ddots & \ddots & \\
     & \ddots & \ddots & t_{n-2,n-1}\\
     0 & & t_{n-1,n} & t_{n-2}+t_{n-1}
    \end{pmatrix},
    \\
D&=\text{diag}(w_1,\ldots,w_n) \qquad \text{and}
 &
 U&:=\begin{pmatrix}
     -u_1, &  \;\phantom{-}u_1, & & & 0\\
     \phantom{-}u_1, & \;-v_1, & \ddots & & \\
     &  \ddots & \;\ddots & \ddots & \\
     & & \;\ddots & - v_{n-2}, & \;\phantom{-}u_{n-1}\\
     0 & & &  \phantom{-}u_{n-1}, & \;-u_{n-1}
    \end{pmatrix}.
\end{align*}
Thus, $U$ and $D$ are of size $n \times n$, $T$ of size  $(n-2) \times (n-2)$ and and $Q$ of size $n \times (n-2)$.\\

These formulae still depend on the unknowns $\lambda_i$ and $\mu_i=\lambda_i \cdot h_i$. 
According to \cite{Rentrop}, the tension parameters $\lambda_i$ can be chosen as uniformly distributed random values in the interval $[4 h_i, 15 h_i]$. \\

Finally, the Lagrange parameter $p$ is still unknown and can be computed by $F(p)^2=S$ with
$$F(p)=\vert\vert p D^{-1} (U-QT^{-1} Q^T -pD^{-2})^{-1} D^{-2}y- p D^{-2} y \vert\vert_2\;.$$
This can be done, for example, by a Newton iteration or interval nesting. 
Thereby, it must be taken into account that the starting value $p^{(0)}=0$ should be avoided since it would lead to vectors $s=0,d=0$ and therefore to a spline $s(x)=0$, see \cite{WGW}.

\section{Monte Carlo results}\label{sec:results}
We simulate the Wilson plaquette gauge action \cite{Wilson}
$$
S(U)=\frac{\beta}{6} \sum_{p} \text{Re} \text{Tr} \{1-U_{p}(t)\}
$$
for gauge group SU(3)
using the Hybrid Monte Carlo algorithm \cite{HMC}.
We let HMC simulations run for varying $\beta$ ($6.03 \leq \beta \leq 6.10$) and the lattice sizes $N_t \times N_s^3$ with $N_t=8$ and $N_s=\{32,40,48\}$.
Taking the gauge configurations of the HMC simulations as initial values for the Wilson flow, we computed the Wilson flow up to $\sqrt{t}/r_0=0.15$ and measured its spatial and temporal energy density. 
Statistical errors and autocorrelation times are computed with the method of
\cite{UWerr}.
Besides, we computed the Polyakov loop susceptibility along the Wilson flow for comparison reasons. 

\subsection{Determination of the critical temperature}
For the determination of the critical temperature, we computed the mean energy difference $\langle \Delta E \rangle$ for all $\beta$ at the flow time $\sqrt{t}/r_0=0.15$ and for each lattice separately.
Then, the data $(\beta, \langle \Delta E \rangle)$ are fitted by exponential smoothing splines $s(\beta)$, see figure \ref{fig:energyDiff}.
Here, the smoothing parameter $S$ is set to $S=2$ and the weights $w_i$ used in the matrix $D$ are set to the statistical errors of $\langle \Delta E \rangle_i$. 
Note that a small variation of $S$ does not change the result; if $S$ is chosen too large, the exponential smoothing spline does not fit each value of $\langle \Delta E \rangle$ in the region of its error $\delta \langle \Delta E \rangle$ .
The maximum of the slope  $s^\prime(\beta)$ of this spline leads to the critical coupling $\beta_c(\Delta E, N_t, N_s)$ which is the value of $\beta$ at which $s^{\prime\prime}(\beta)=0$ holds. 
The errors $\delta \beta_c(\Delta E, N_t, N_s)$ are computed by a small variation of the single values of $\langle \Delta E \rangle$ using the Gaussian error propagation law:
$$\delta \beta_c^2 = \sum_{k=1}^{n_\beta} \biggl(\frac{\partial \beta_c}{\partial \langle \Delta E \rangle_k}\biggr)^2 \cdot \langle \Delta E \rangle_k^2 .$$
The partial derivatives $\partial \beta_c / \partial \langle \Delta E \rangle_k$ are computed using the symmetric difference quotient with perturbed value $\Delta E_k$; we use 10 percent of the error of $\langle \Delta E \rangle_k$.  

\begin{figure}\begin{center}
    \includegraphics[width=10cm]{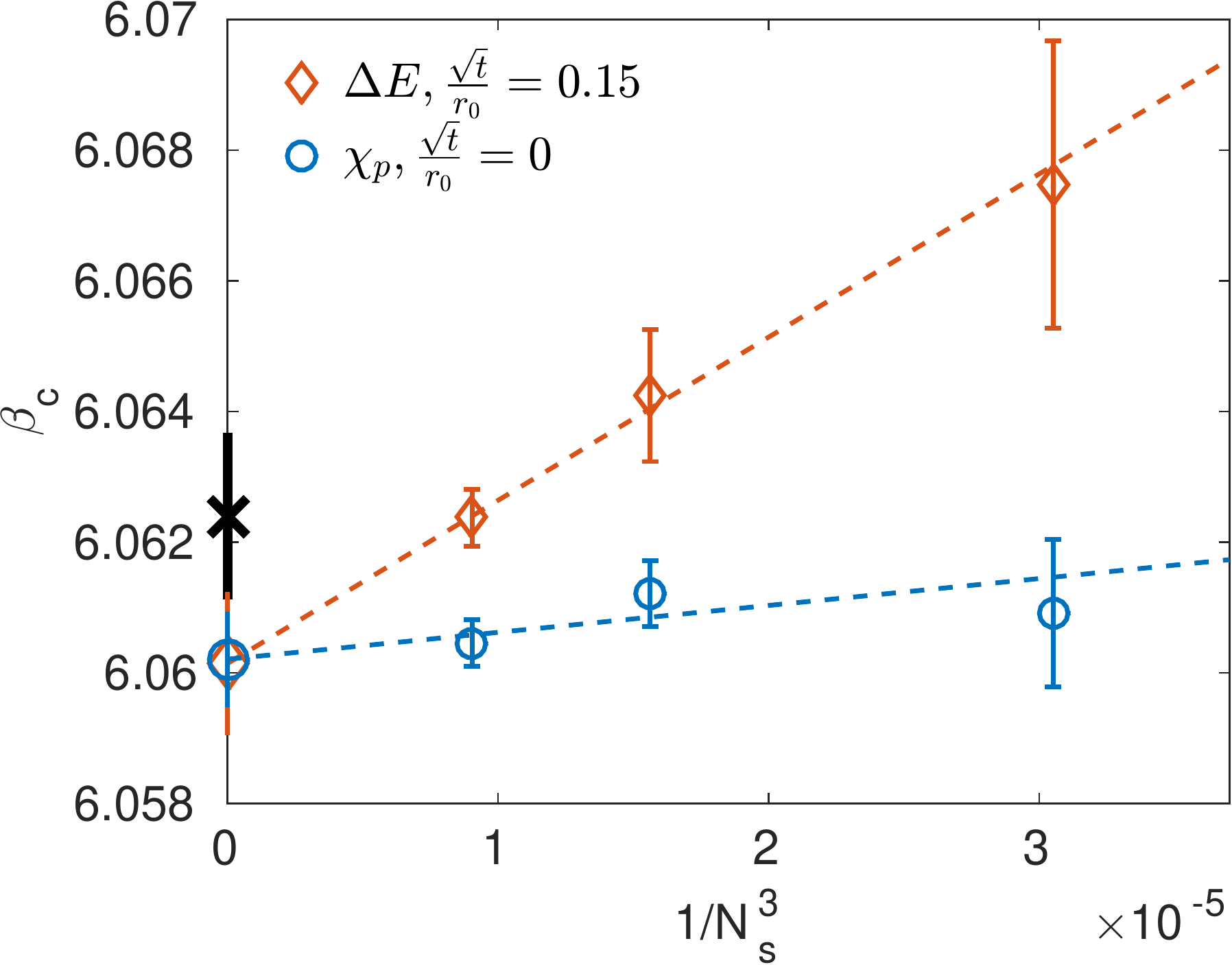}
    \caption{The critical couplings $\beta_c(N_s=\{32,40,48\})$ for the different lattice sizes and $N_t=8$ are shown for the energy difference method (red diamonds) and the standard Polyakov loop susceptibility approach (blue circles). The extrapolated values $\beta_c(\Delta E, N_s=\infty)$ and $\beta_c(\chi_p, N_s=\infty)$ of both methods coincide. 
    The black cross represents the reference value $\beta_c(N_t=8, N_s=\infty)=6.0624(12)$ from \cite{betacNt8}.}
    \label{fig:res}
\end{center} \end{figure}

At the end, a linear extrapolation in $1/N_s^3$ leads to the final value $\beta_c(\Delta E, N_t=8, N_s=\infty) = 6.0601(11)$ which is shown in figure \ref{fig:res}. 
It agrees within errors with the value $6.0624(12)$ computed in \cite{betacNt8} and with the newest result $6.06239(38)$ of Ref.~\cite{betacNt8new}. 
Using equation \eqref{eq:ln:a:r0} this result translates to $T_c r_0 = 0.7426(14)(37)$, where the first error is the uncertainty in $\beta_c$ and the second one is a $0.5 \%$ error for $r_0/a$ computed from equation \eqref{eq:ln:a:r0}. 
For comparison reasons, we determined the critical coupling from our data with the standard Polyakov loop susceptibility approach described in paragraph \ref{subsec:PLS} and shown in figure \ref{fig:polyakov}. 
Also this method leads to a critical coupling  $\beta_c(\chi_p, N_t=8, \infty)= 6.0602(7)$ close to the value given in the literature \cite{betacNt8}.
Both methods lead to very consistent values of the critical coupling and also the size of the errors are of the same magnitude.  

\subsection{Statistical errors at different flow times}\label{subsection:stat_err}
\begin{figure}\begin{center}
    \includegraphics[width=7cm]{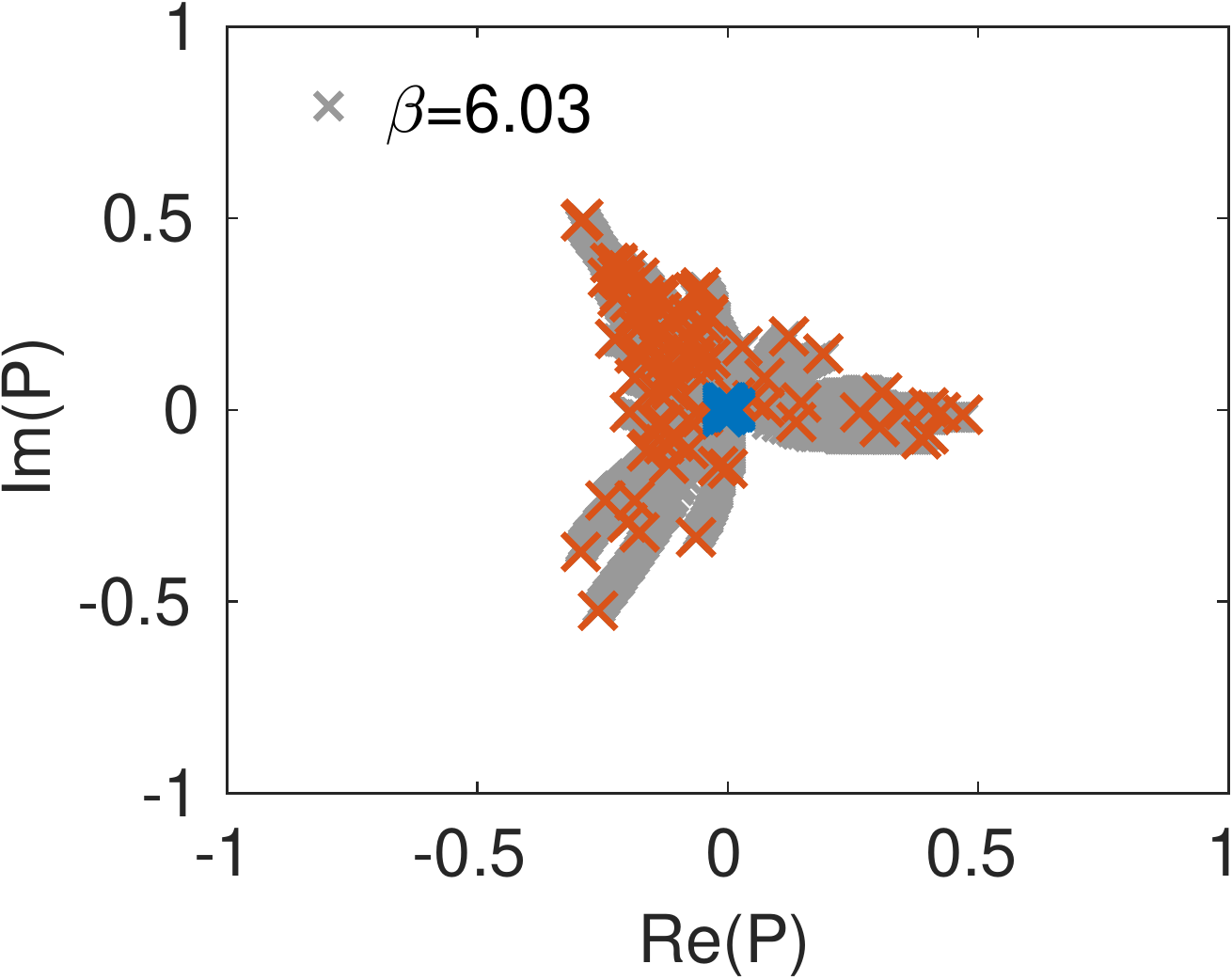}
    \hspace{0.5cm}
    \includegraphics[width=7cm]{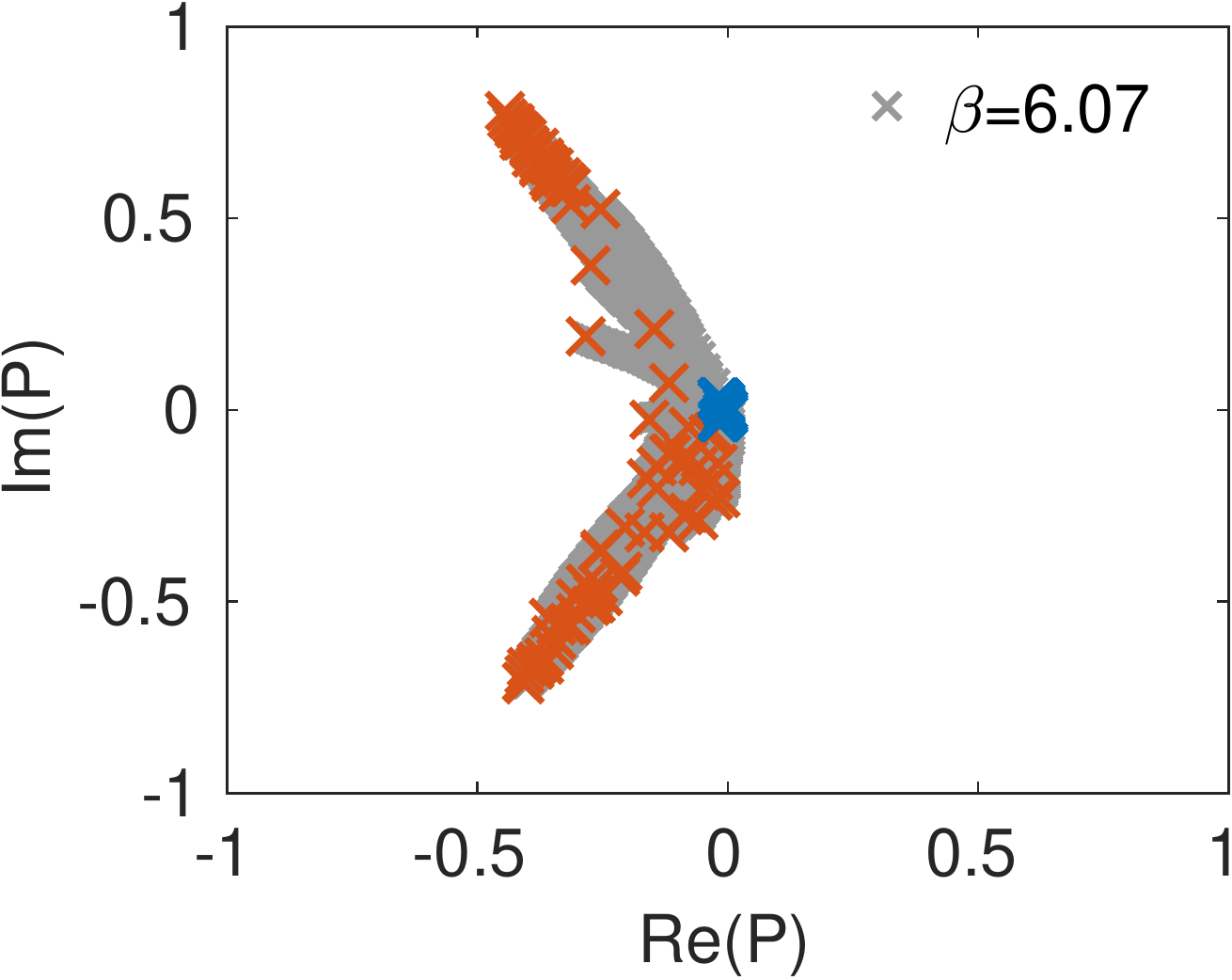}
    \caption{The value of the Polyakov loop along the Wilson flow from $\sqrt{t}/r_0=0$ up to $\sqrt{t}/r_0=0.708$ for $\beta < \beta_c$ (left) and $\beta > \beta_c$ (right) for 100 ensembles of gauge configurations. The value of the Polyakov loop for flow time $\sqrt{t}/r_0=0$ (blue) is close to the origin, departs during the Wilson flow (gray) and ends at $\sqrt{t}/r_0=0.708$ (red).}
    \label{fig:polyakovLoop}
\end{center} \end{figure}

We investigated the statistical errors of both methods at different flow times. 
For the Polyakov loop susceptibility, the relative statistical errors do not change during the flow. The Wilson flow has just an effect on its absolute value, see figure \ref{fig:polyakovLoop}.
Therefore, it would be no advantage to combine the Wilson flow and the Polyakov loop susceptibility. 
On the other hand, the statistical errors for the energy difference method have a minimum at $\sqrt{t}/r_0=0.15$. 
We checked that this flow time is large enough such that cut-off effects (measured by comparing different definitions of the energy density) are small and therefore the best choice for the computation of the critical temperature. 

\begin{figure}[h]\begin{center}
    \includegraphics[width=7cm]{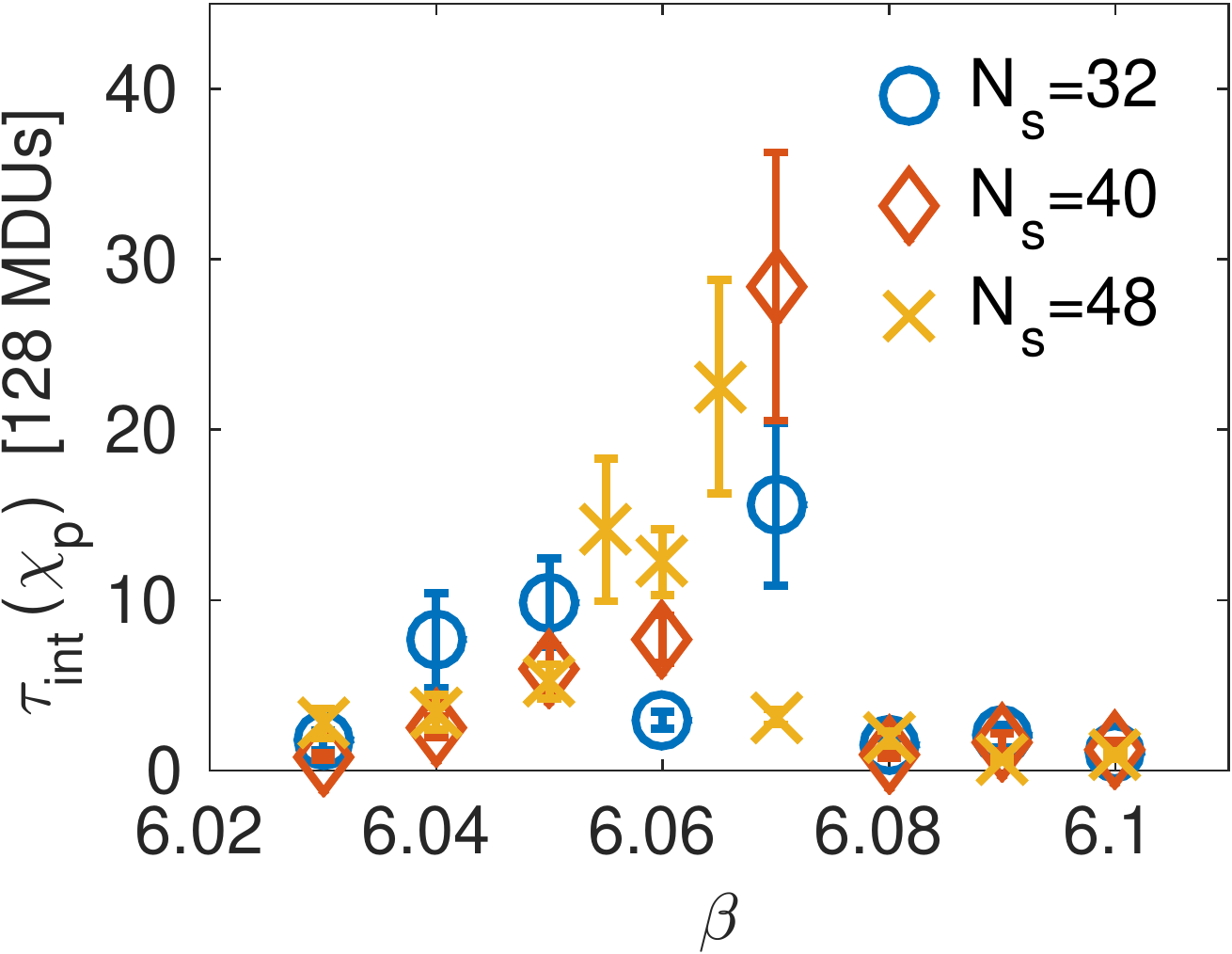}
     \hspace{0.5cm}
    \includegraphics[width=7cm]{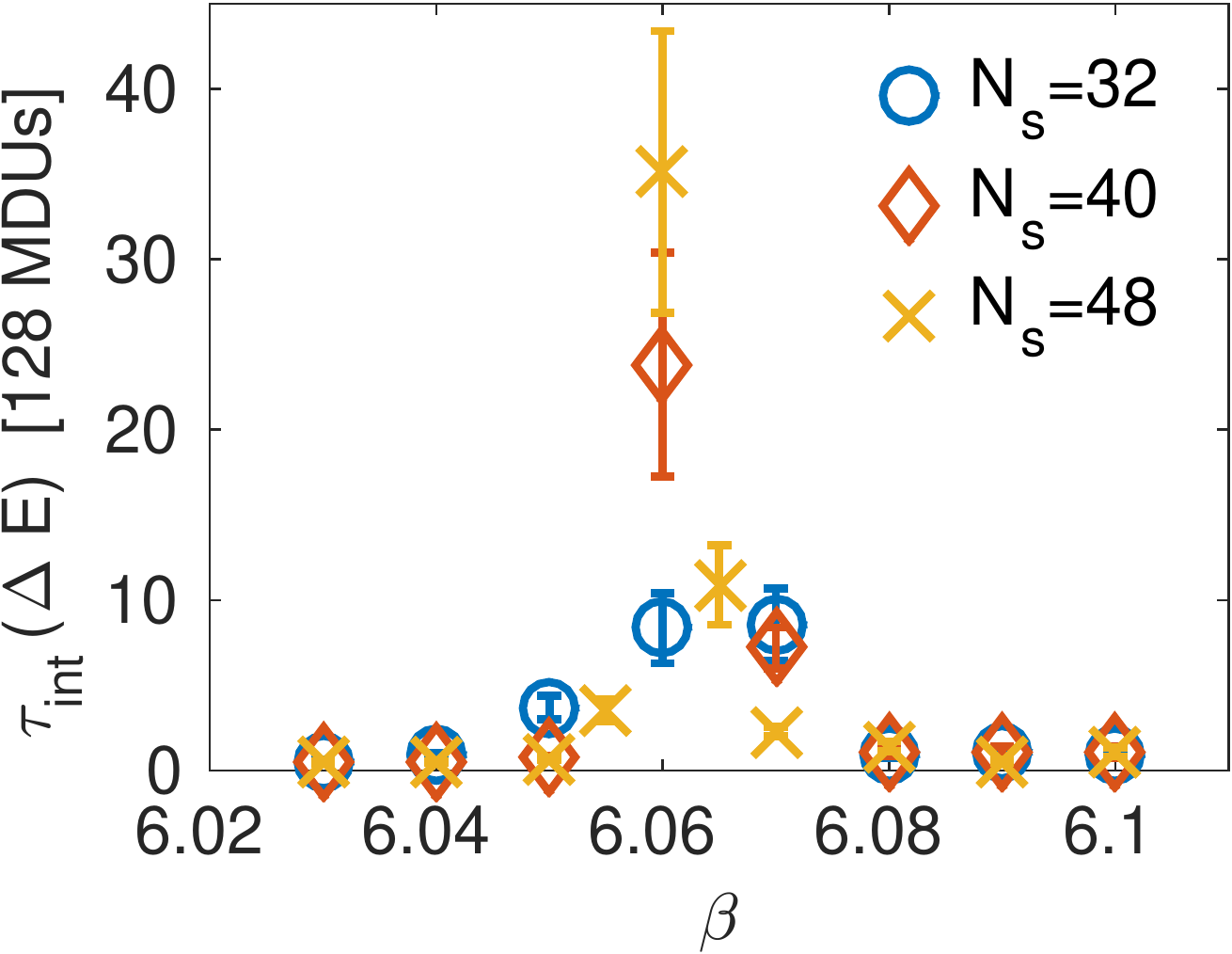}
    \caption{Integrated autocorrelation times of the Polyakov loop susceptibility (left) and the energy difference (right) in units of 128 MDUs. }
    \label{fig:tauint}
  \end{center} \end{figure}

The statistical errors already include autocorrelation effects shown in figure \ref{fig:tauint} which are quite large close to the critical coupling.
For this reason, long simulations have to be run for these values of the coupling constant $\beta$ such that the number of independent configurations is large enough. 
Surprisingly, the number of independent configurations for the energy difference at flow time $\sqrt{t}/r_0=0.15$ is for almost all values of $\beta$ (except at $\beta_c$) larger than the one of the Polyakov loop susceptibility at $\sqrt{t}/r_0=0$, mentioned in table \ref{tab:indcount}. 
This means, the number of configurations needed for the energy difference method is smaller than for standard approach.  
 
 \begin{small}
  \begin{table}
  \begin{center}
  \begin{tabular}{|l||c|c||c|c||c|c|}
  \hline 
   & \multicolumn{2}{|c||}{$N_s=32$} & \multicolumn{2}{|c||}{$N_s=40$} & \multicolumn{2}{|c|}{$N_s=48$}\\
   \hline
   & $\chi_p$ & $\Delta E$  & $\chi_p$ & $\Delta E$ & $\chi_p$ & $\Delta E$\\
   \hline 
   & & & & & & \\[-2.9ex] 
   \hline
   $\beta=6.03$ & 62(20) &   208(39) & 
   166(38) & 287(34) & 
   72(22) & 426(43)\\
   $\beta=6.04$ & 36(13) & 312(56) & 
   135(32) & 683(73) & 
   74(21) & 508(45)\\
   $\beta=6.05$ & 101(26) & 269(49) & 
   134(31) &  1034(112) & 
   238(46) & 1784(139)\\
   $\beta=6.055$ & - & - & - & - & 70(20) & 280(50)\\
   $\beta=6.06$ & 339(57) & 119(29) & 
   265(48) & 86(23) & 
   380(59) & 132(31)\\
   $\beta=6.065$ & - & - & - & - & 84(23) & 173(37)\\
   $\beta=6.07$ & 63(19) & 116(29) & 
   85(23) &  337(55)& 
   540(74) &  746(89) \\
   $\beta=6.08$ & 134(33) & 215(46) & 236(48) & 198(44) & 145(34) & 193(40)\\
   $\beta=6.09$ & 72(23) & 145(35) & 70(22) & 105(29) & 246(47) & 279(44)\\
   $\beta=6.10$ & 195(45) & 198(42) & 51(18) & 61(19) & 168(40) & 143(34)\\    
   \hline         
  \end{tabular}
  \end{center}
\caption{Number of independent configurations. } 
\label{tab:indcount}
 \end{table}
 \end{small}

 \subsection{Conclusion and outlook}\label{subsec:CO}
 For pure gauge theory, it is possible to detect the critical coupling $\beta_c$ via the energy difference of the Wilson flow.
 Our results agree with the standard method as well as with the reference value given in \cite{Boyd}, \cite{betacNt8} and \cite{betacNt8new}.
 For the energy difference method, the Wilson flow has to be computed in addition to the HMC as it leads to a reduction of the statistical error. 
 This is a disadvantage compared to the standard approach but just a small one, since the energy difference can be evaluated at small flow times.
 For most values of $\beta$ the statistical errors of the energy difference method are smaller than for the standard approach such that the simulation has to be run for less configurations to reach the same amount of independent configurations.\\
 
Moreover, our results are promising for simulations with fermions.  
Our method can be used as alternative to the expensive computation of the chiral susceptibility, which is usually taken in this case. 
Our technique based on the exponential smoothing spline is also suitable for other applications, as it allows to fit a smooth function through a data set and compute its derivatives.
Additionally, the integration of the Wilson flow can be improved by using better integrators. 
For example, adaptive step size methods developed in \cite{Fritsch:Ramos} and \cite{WG} reduce at the same time the computational cost and are able to control the statistical errors.


\acknowledgments
We thank M. L\"uscher for sharing with us his code for simulations of the Yang-Mills SU(3) theory. 
We thank Jacob Finkenrath and Bj\"orn Leder for their contributions in an early stage of this work.
This work is part of project B5 within the SFB/Transregio 55
{\em Hadron Physics from Lattice QCD} funded by DFG (Deutsche Forschungsgemeinschaft).





\end{document}